# Retrofitting Security into a Web-Based Information System

David Bettencourt da Cruz, Bernhard Rumpe, Guido Wimmel
Software & Systems Engineering, Technische Universität München
85748 Munich/Garching, Germany

This paper reports on an incremental method that allows adding security mechanisms to an existing, but insecure system, such as a prototype or a legacy system. The incremental method is presented and as a showcase its application is demonstrated at the example of a Web-based information system.

## 1 Introduction

Security is an extremely important issue in the development of distributed systems. This applies in particular to Web-based systems, which communicate over an open network. Failures of security mechanisms may cause very high damage with financial and legal implications. Security concerns, both on the part of enterprises and consumers, are one of the major reasons why new technologies such as E-commerce or E-government are used very reluctantly.

Developing security-critical systems is very difficult. Security is a complex non-functional requirement affecting all parts of a system at all levels of detail. To secure a system, merely adding mechanisms such as cryptography in some places is not sufficient. Whether a system is secure depends crucially on the complex interplay of its components and the assumptions about its environment. A single weakness can compromise the security of the entire system.

Furthermore, many systems are developed initially without security in mind. Reasons for that are that they were designed for a secure environment such as a local network, that existing legacy systems are to be adapted, or because they were first developed as a functional prototype. Retrofitting security into an existing system is generally believed to be extremely hard to achieve, and it is in effect often advised against doing so at all. In this article we report on the experiences of a Java project where exactly this retrofitting was done after developing initial prototypes.

The RAC system is an Internet information system based on the "push" principle: information is presented to the user on a client application ("pushlet") and updated when necessary, without the user having to explicitly check for such updates. The server regularly or on demand contacts the client for updates. The RAC system was initially developed as a prototype without security functionality as its focus was targeting to be production companies' internal information systems.

In this paper, we describe a method to carry out a security analysis of an existing system and to introduce appropriate mechanisms to achieve high trustworthiness. Our method is demonstrated at the example of the RAC system. It is based on a combination of an evolutionary approach and method suggested in [1]. We comment on ex-periences and difficulties in adding security to an existing system, in particular in the context of Web-based Java applications.

**Related Work.** The consideration of additional or changed requirements within the lifetime of a system is one of the main aims of iterative processes, such as Boehm's Spiral Model [4]. Few works are available on the integration of security aspects into the development process. In [1], Eckert suggests a top-down approach, which we used as a basis for our work. A mapping of ITSEC security requirements to development activities in the German V-Model 97 is given in [5]. [3] describes a lifecycle process based on the Evaluation Assurance Requirements of the Common Criteria for Security Evaluation, at the example of a payment system. These processes are mainly tailored to the development of new systems. Security aspects of distributed Java applications are covered in detail in [6], but methodical guidance is missing there.

## 2 The Web-Based Information System RAC

The RAC system is an experimental prototype serving a variety of issues. It is a Web-based information retrieval system that updates its information automatically by pushing new information to its clients. Therefore the presented information is always up to date no matter whether the information changes within seconds (such as in stock information systems) or within minutes or hours (such as e.g. temperature values). Another pleasant effect of the pushing mechanism is that the system has a very efficient communication (no polling needed), which can even be used over low-bandwidth communication lines. For more information about the RAC system, see [9].

## 3 Method for Introducing Security Features

Early in the development, it was decided that the RAC system is to be developed in increments. Also based on our experiences in building similar systems it was decided to build an efficient feasibility prototype without any security mechanisms. Instead, any security considerations should be retrofitted into the existing system in a later increment. We were well aware that this might make it necessary to refactor parts of the code.

The method used for adding security to the RAC system was based on existing methodologies for developing new security-critical systems ([1], [3]). The main steps taken here and the differences and difficulties found upon retrofitting them into an existing system are sketched in the rest of the section.

**Threat Analysis.** During the threat analysis each and every possible threat to the system has to be documented (threats are situations or events that may lead to unauthorized access, destruction, disclosure or modification of data, or to denial of service). Obviously, only threats that have been identified at this stage can later on be considered to be countered. Therefore, it is important that the threat analysis is as complete as possible. Hence it is crucial to use a systematic approach to identify the threats. In case of the RAC-System, threat trees [7] were used. The root of a threat tree consists of all possible threats to a system. Its successor nodes correspond to more fine-grained threat classes (which together make up all possible threats), and the leafs consist of single threats or very small related groups of threats. There is some degree of freedom in how the threats are decomposed, as long as no threats are lost during the process.

**Threat Classification.** After having completed the search for the threats, the resulting threats are classified (1) by an estimate of the potential *damage* caused if the threat can be realized, and (2) by an estimate of the *effort* it would take an attacker to realize the threat. To allow for a systematic risk analysis, damage and effort are measured in a quantitative metrics, depending on the use of the program. In programs with a commercial use, money is mostly a good scale to choose. Other appropriate metrics are necessary time, hardware or knowledge, which can in turn again be represented by money.

In case of the RAC system both, the threat analysis and the threat classification, were greatly facilitated by the available increments. The necessary work for the threat analysis could be reduced from finding threats that might appear in future to finding threats that could actually be encountered in the running version. The classification of identified threats was also easier, since the available components immediately gave an idea of the values for potential damage and effort and if they didn't it was possible to simply try out and check the results.

**Finding Countermeasures.** The next step to take is to find countermeasures against each of the identified threats independently of how they have been classified. Fortunately, quite a number of standard security techniques, patterns and concepts [1,2] that provide countermeasures against most of the major threats already exist. At this step the countermeasures found against the threats do not yet have to be worked out in great detail. Providing a basic idea is usually enough to be able to classify them in the next step.

**Classification of the Countermeasures.** The classification of the countermeasures is carried out in a way similar to the way used in threat analysis. For this classification, the effort needed to realize a countermeasure is estimated. Again the classification is based on a metrics and thus allows comparison. Ideally the metrics used is the same as the one used for threat classification, since that would ease the following combination of all classifications.

When retrofitting security into a system, specifying and assessing countermeasures is more difficult, as the design of the existing system must be taken into account. If the countermeasure against a threat cannot be assigned to a small, modular or easily separable part of the program, it becomes much harder to retrofit. Therefore it is very important to classify the countermeasures with great care in order to implement the correct security features in the next increment.

**Combining the Classifications.** Finally the results from all the classifications are combined to identify which threats should be dealt with. This largely depends on the available budget and the level of security that must be reached. This procedure will result in the functional security requirements, which then can be used as if they were traditional requirements and implemented. As a specialty, we found that these so called "requirements" usually go deep into design activities and thus combine analy-sis and design phase. In this respect, we deviate from the strictly sequential approach suggested in [1]. Postponing the decision which threats should be countered to the point after the assessment of corresponding countermeasures adds a little overhead. But it is easier in our case as there is already an implemented system available and it leads to a more effective selection of security measures within the given time/budget.

More detailed information on how this approach was applied to the RAC system can be found in [9].

## 4 Discussion

In our approach, the choice of the countermeasures that should be implemented is based on both an assessment of the threats and of the countermeasures. This was only possible with reasonable time and effort because an implementation and automated tests were already available. Based on this information, the most effective security functions given particular time/budget constraints for their implementation can be selected. The existing tests could be re-used to verify that the program's functionality has not been affected, and additional tests for the security functionality could be added. Finally, it is important that the existing software is well documented. Otherwise the threat analysis becomes hard to carry out.

We believe that Java is to date the most appropriate architecture for information systems where security might become important after some iterations, because of its built-in security features. However, there are also problems. Firstly, the Java Sandbox does not implement the principle of complete mediation [8]. It usually only checks for correct access when a protected object is created. Further use of the object works independently of the Sandbox, thus enabling attackers to gain access to the object through the program. Secondly, if the program is not strictly modularized, the Sandbox becomes much harder to introduce, and restructuring might become necessary to be able to use the access modifiers (private, protected, public).

If these points are considered, a certain level of security can be actually retrofitted into existing software without great overhead.